\pdfoutput=1

\documentclass{article}
\usepackage[utf8]{inputenc}
\usepackage[T1]{fontenc}
\usepackage[sfdefault,scaled=.85]{FiraSans}
\usepackage{newtxsf}
\usepackage[protrusion=true, expansion]{microtype}
\usepackage[margin=18mm, bottom=23mm]{geometry}
\usepackage{graphicx}
\usepackage{amssymb}
\usepackage{amsmath}
\usepackage[hidelinks,
  pdfauthor={},
  pdftitle={},
  pdfkeywords={}
]{hyperref}
\usepackage{bookmark}

\begin{document}

\noindent\hrule

\medskip

\begin{center}

  {\Large\bf Boron nanotube structure explored by evolutionary computations}

  \bigskip
  
  {
    \large
    \href{https://orcid.org/0000-0002-6071-9389}{Tomasz Tarkowski}\,\textsuperscript{1}
    and
    \href{https://orcid.org/0000-0002-0518-844X}{Nevill Gonzalez Szwacki}\,\textsuperscript{2}*
  }

\end{center}
  
\medskip

{
  \small\it
  \noindent\textsuperscript{1} Institute of Theoretical Physics, Faculty of Physics, University of Warsaw, Pasteura 5, PL-02093 Warsaw, Poland; tomasz.tarkowski@fuw.edu.pl \\
  \textsuperscript{2} Institute of Experimental Physics, Faculty of Physics, University of Warsaw, Pasteura 5, PL-02093 Warsaw, Poland; gonz@fuw.edu.pl \\
  * Correspondence: gonz@fuw.edu.pl; Tel.: +48-22-5532797
}

\bigskip\bigskip

\begin{abstract}
  \noindent In this work, we explore the structure of single-wall boron nanotubes with large diameters (about 21~{\AA}) and a broad range of surface densities of atoms. The computations are done using an evolutionary approach combined with a nearest neighbors model Hamiltonian. For the most stable nanotubes, the number of 5-coordinated boron atoms is about $63\%$ of the total number of atoms forming the nanotubes, whereas about $11\%$ are boron vacancies. For hole densities smaller than about 0.22, the boron nanotubes exhibit randomly distributed hexagonal holes and are more stable than a flat stripe structure and a quasi-flat B$_{36}$ cluster. For larger hole densities ($> 0.22$) the boron nanotubes resemble porous tubular structures with hole sizes that depend on the surface densities of boron atoms.

  \medskip


  \begin{flushright}

  {\scriptsize Copyright (c) 2022 Tomasz Tarkowski and Nevill Gonzalez Szwacki. License: CC BY 4.0 (\url{https://creativecommons.org/licenses/by/4.0/}).}
  \end{flushright}
\end{abstract}

\medskip

\noindent\hrule

\bigskip

\section{Introduction}
Boron nanotubes (BNTs) have been synthesized for the first time in 2004 \cite{Ciuparu2004}. A magnesium-substituted mesoporous silica template (Mg-MCM-41) was applied by Ciuparu \textit{et al}.~\cite{Ciuparu2004} to prepare pure boron single-wall nanotubes at $870{ }^{\circ} \mathrm{C}$ with uniform diameters ($36 \pm 1$~{\AA}) using the mixture of $\mathrm{BCl}_3$ and $\mathrm{H}_2$ as gas sources. The authors attributed the Raman peaks at $210 \mathrm{~cm}^{-1}$ and between $300$ and $500 \mathrm{~cm}^{-1}$ to typical tubular structures, where the first ($210 \mathrm{~cm}^{-1}$) corresponds to the characteristic radial breathing mode. In 2010 Liu \textit{et al}.~\cite{Liu2010} reported the first large-scale fabrication of single crystalline multilayered BNTs using boron $(99.99 \%)$ and boron oxide powders $(99.99 \%)$ as source materials. The as-synthesized BNTs had lengths of several micrometers and diameters in a range from 10 to 40 $\mathrm{~nm}$. The nanotubes were cataloged as multilayered single crystalline BNTs with an interlayer spacing of about $3.2$~{\AA}. Moreover, these nanotubes were experimentally proven to have metallic properties regardless of their chirality. The metallic behavior of BNTs
makes them attractive in the design of novel electronic nanodevices, such as field-effect transistors, light-emitting diodes, and field-emission displays or for photosensitive device applications \cite{Tian2010,Tian2019}.

The experimental studies were preceded by several theoretical investigations, most often based on density functional theory (DFT). Quasiplanar \cite{Boustani1997a,Boustani1997b}, tubular \cite{Boustani1997c,Gindulyte1998}, convex, and spherical \cite{Boustani1999} boron clusters have been computationally explored. Moreover, the existence of quasiplanar boron clusters \cite{Kiran2005,Piazza2014} implies that boron fullerenes \cite{GonzalezSzwacki2007a,Zhai2014} and caped BNTs \cite{GonzalezSzwacki2007b, Chernozatonskii2008} may exist because larger in size quasiplanar or planar clusters will tend to remove dangling edge bonds by forming closed tubular or polyhedral structures.

Although some reports affirm that BNTs with diameters smaller than 1.7~nm \cite{Singh2008,Yang2008} or 2~nm \cite{Tang2010} show semiconducting behavior due to the band opening at the Fermi level through curvature-induced out-of-plane buckling of certain atoms, later calculations based on second-order Møller–Plesset perturbation theory \cite{GonzalezSzwacki2010} and dispersion-corrected DFT calculations \cite{Gunasinghe2011} showed that the surface buckling is more likely an artifact of standard DFT approaches. 

Thus, BNTs are found to be metallic and independent of diameter and chirality (armchair or zigzag) \cite{Kunstmann2014}. Moreover, there have been a lot of morphologies of BNTs in contrast to only one morphology of carbon nanotubes (CNTs) \cite{Caliskan2022}. A more detailed description of the structure and properties of 1D-boron structures can be found in Refs.~\cite{Kondo2017,Tian2019}.

Crystal structure prediction (CSP) is a long-standing challenge in physical and materials sciences. Several methods have been applied for materials design \cite{Dieb2019,Paszkowicz2009}. Evolutionary algorithms, such as genetic algorithms (GAs) \cite{Paszkowicz2009} that use human evolution mechanisms (such as crossover and mutation), have been used, for instance,  to obtain the most stable structures of prototype nanotubes composed of particles interacting through a Lennard-Jones potential \cite{Davies2007}.

Boron two-dimensional structures \cite{Tang2007} conceptually form foundations for BNTs by ``cutting'' adequate stripes and ``gluing'' them along the edges. Our previously developed methodology based on the application of the floating-point representation is not the only possible strategy for the CSP of boron nanostructures \cite{Tarkowski2022}. Alternatively, one can use binary (for monoatomic materials) or integer (for alloys) representations with a fixed crystal lattice. To obtain the total energy of the system, several approaches can be used. In our previous work, we used a genetic algorithm \cite{tarkowski_tomasz_2022_6484743} combined with DFT to predict the structure of boron nanowires \cite{Tarkowski2022}. In this work, we will use a GA combined with a DFT-based nearest neighbors model Hamiltonian to predict the structure of BNTs. We assume that the nanotubes have a perfect cylindrical shape and their structure is closely related to one-atom-thick sheets of boron atoms arranged on a hexagonal lattice \cite{Tang2007}.

\section{Computational approach and results of simulations}
\subsection{Unit cell definition}
The information about the position of atoms and vacancies (``holes'') on a hexagonal lattice wrapped cylindrically around a nanotube axis is encoded using a binary representation. The type of considered nanotubes is limited to an armchair $(n,n)$ chirality as described in Ref.~\cite{Gindulyte1998}.  A cylindrical coordinate system can be conveniently used for the nanotube description. It is assumed that the nanotube has periodic boundary conditions (PBCs) in the $z$ direction. The unit cell size is described with two numbers $n_\phi , n_z \in \mathbb{N}_+$, where $n_\phi > 1$. The lattice constant (interatomic distance) of the unwrapped hexagonal layer is equal to $a$ (to plot the BNTs, we have assumed $a=1.675$~{\AA}). The positions of atoms and vacancies are not encoded in the genotype but are specified by the following position vectors:
\begin{equation}
\vec{r}_i = \left(
\frac{a \sqrt{6}}{4 \sqrt{1 - \cos ( \pi / n_\phi)}} ,
\frac{\pi}{n_\phi} \cdot \left\lfloor \frac{i}{n_z} \right\rfloor ,
(i \bmod n_z) \cdot a + \left\{
\begin{array}{ll}
  0 & {\rm if}\ \lfloor i / n_z \rfloor \, {\rm is\ even} \\
  a / 2 & {\rm otherwise}
\end{array} 
\right.
\right)_{(\rho , \phi , z)}
.\end{equation}
The use of binary representation, $g = (x_i)_{i = 0}^{c - 1} \in \mathbb{B}^c$, where $c = 2 n_\phi n_z$, justifies the way the vectors $\vec{r}_i$ are indexed (more details about indices are given in Appendix~A). It also implies the number of atoms in the unit cell equal to $N_{\rm a} = \# \{ i \in \iota_c \mid x_i \}$ and the number of vacancies equal to $N_{\rm v} = c - N_{\rm a}$.

The unit cell of the BNT is defined in terms of predicate theory. Predicate $Q \equiv Q_0 \wedge Q_1 \wedge Q_2$, where predicates $Q_0$, $Q_1$, and $Q_2$ state that atoms within the unit cell are connected, that there is an atomic neighborhood at the boundary of two unit cells along the nanotube axis, and that there is an atomic neighborhood at the unit cell boundary perpendicular to the circumference of the nanotube, respectively. The predicate Q is sufficient to ensure the nanotube connectivity and its closure at the circumference (see also Appendix~A for a more precise formulation of the predicate Q). It does not restrict any possible valid result of structures that are equivalent by translational or rotational symmetry.

\subsection{Fitness function}
The fitness function values are calculated using a nearest neighbors model Hamiltonian. The optimization task consists in finding such configuration of atoms and vacancies which maximizes the fitness function---the binding energy per atom, $E_{\rm b}$, calculated within the nearest neighbors model Hamiltonian. In this model, $E_{\rm b}$ can be defined as:
\begin{equation}
  E_{\rm b} (n_1, n_2, n_3, n_4, n_5, n_6) = \frac{1}{N_{\textnormal{a}}} \sum_{i=1}^6 n_i e_i
,\end{equation}
where $n_i$ is the number of atoms in the unit cell with $i$ nearest neighbors, while $e_i$ is the $E_{\rm b}$ for a structure consisting of atoms having only $i$ nearest neighbors (e.g., two-atom molecule, single atomic chain, or a honeycomb structure, \emph{hc}, for $i$ equal to $1$, $2$ or $3$, respectively). Obviously, the total number of atoms in the unit cell $N_{\rm a} = \sum_{i = 1}^6 n_i$. The values of $e_i$ are taken from our previous work \cite{Tarkowski2021} and they are equal to $e_1 = 1{.}7803$, $e_2 = 5{.}1787$, $e_3 = 5{.}6504$, $e_4 = 6{.}2522$, $e_5 = 6{.}5718$, and $e_6 = 6{.}5116$ in units of ${\rm eV} / {\rm atom}$. Since our model Hamiltonian includes only nearest neighbors interactions and was developed for the case of a one-atom-thick layer of boron atoms \cite{Tarkowski2018, Tarkowski2021}, it does not take into account the curvature of the BNT. Therefore, the $e_i$ parameters can be applied to nanotubes with large radii, where the effect of curvature on the properties of the nanotube is expected to be negligible.

In the case of carbon, the elastic properties of the nanotubes do not depend on the diameter already at about 18~{\AA} \cite{Hernandez1999}. The same is true for electronic properties such as the energy gap \cite{Umari2012}. Making the assumption that the pure boron case situation is the same, $n_\phi = 23$  is taken which corresponds to $\rho \approx 10{.}628\,\text{\AA}$. This gives the size of the potential solution space (described by a general formula of $2^{2 n_\phi n_z}$) equal to $2^{46 n_z}$.

The Bernoulli ${\rm B}(1, 0{.}5)$ probability distribution [$B(n,p)$ is a discrete probability distribution, where the Bernoulli random variable $n$ can have only 0 or 1 as the outcome values, $p$ is the probability of success ($n=1$), and $1 - p$ is the probability of failure ($n=0$)] was chosen for the creation of the first generation of nanotubes, rejecting genotypes not satisfying the predicate $Q$. A one-point recombination with bit-flipping mutation ($p = 1 / c$) applied stochastically with probabilities equal to $p_{\rm r} = p_{\rm m} = 0{.}5$ is employed during the evolution. 

The generation size $\mu$ is set to $4 n_\phi n_z$ while parent multiset size $2k = 2 n_\phi n_z$. Stochastic universal sampling (SUS) with linear ranking selection ($s = 2$) is used for the parent selection and selection to the next generation mechanisms. The genetic search is terminated after reaching a \emph{plateau} for the fitness function or, more precisely, when after $100$ generations the fitness function maximum has not improved by $\Delta E = 1\,{\rm meV}$.

Some of the obtained nanotubes can be equivalent through rotations around the $z$-axis and translations along this axis. There is a relationship between natural numbers and the above-defined nanotubes. Encoding $g \in \mathbb{B}^{2 n_\phi n_z}$ corresponds to certain number $n = \sum_{i=0}^{2 n_\phi n_z -1} 2^i x_i$, where $\mathbb{B} \ni {\rm true} \equiv 1 \in \mathbb{N}$ and $\mathbb{B} \ni {\rm false} \equiv 0 \in \mathbb{N}$. There are classes of abstractions of natural numbers regarding rotations around the $z$-axis (nanotube axis) and translations along this axis, which form equivariant nanotubes. Rotations and translations in the set $\iota_{2 n_\phi n_z}$ can be defined as follows:
\begin{align}
  R_{n_\phi , n_z}^{\Delta n_\phi} (n) & = n', \, x_{(i + 2 \Delta n_\phi n_z ) \bmod 2 n_\phi n_z}' = x_i ,\\
  T_{n_\phi , n_z}^{\Delta n_z} (n) & = n', \, x_{\left( (i + \Delta n_z) \bmod n_z \right) + \lfloor i / n_z \rfloor n_z}' = x_i ,
\end{align}
where $\Delta n_\phi , \Delta n_z \in \mathbb{Z}$. The following identities hold:
\begin{align}
  R_{n_\phi , n_z}^{n_\phi} & = {\rm id} ,\\
  T_{n_\phi , n_z}^{n_z} & = {\rm id} ,\\
  R_{n_\phi , n_z}^{\Delta n_\phi} & = \left( R_{n_\phi , n_z}^{1} \right)^{\Delta n_\phi} ,\\
  T_{n_\phi , n_z}^{\Delta n_z} & = \left( T_{n_\phi , n_z}^{1} \right)^{\Delta n_z} ,\\
  R_{n_\phi , n_z}^{-1} & = R_{n_\phi , n_z}^{n_\phi - 1} ,\\
  T_{n_\phi , n_z}^{-1} & = T_{n_\phi , n_z}^{n_z - 1}.
\end{align}
It may also turn out that the nanotube obtained as a result of evolution has a shorter elementary unit cell along $\hat{e}_z$ than it would result from the $n_z$ value. This occurs if, and only if, the maximum divisor $d_{n_z}$  of the number $n_z$, for which the condition
\begin{equation}
  T_{n_\phi , n_z}^{n_z / d_{n_z}} (n) = n
  \label{eq:dnz}
,\end{equation}
is satisfied for $n$ representing a given nanotube, is greater than $1$. If maximum divisor $d_{n_\phi}$ of $n_\phi$, for which analogous condition
\begin{equation}
  R_{n_\phi , n_z}^{n_\phi / d_{n_\phi}} (n) = n
  \label{eq:dnphi}
\end{equation}
is met, is greater than $1$ then the nanotube has a nontrivial rotational symmetry about the axis of the tube. The values of $d_{n_z}$ and $d_{n_\phi}$ are assumed to denote the maximum values satisfying conditions given by Eqs. \ref{eq:dnz} and \ref{eq:dnphi}, respectively. 

\subsection{Results of simulations}

The most stable evolutionary obtained nanotubes are labeled as $\tau_{n_\phi , n_z}$. The computational complexity for a given value $n_\phi$ grows exponentially with unit cell length $n_z$. For $n_\phi = 23$, under the assumption that a single structure is proceeded for $10\,\mu{\rm s}$ even for $n_z = 2$ with sequential computations, the time five orders of magnitude longer than the Universe age \cite{PlanckCollaboration2020} is needed to check all possible potential solutions. The application of a genetic algorithm is therefore advisable.

\begin{figure}
  \centering
  \begingroup
  \setlength{\tabcolsep}{2pt}
  \footnotesize
  \begin{tabular}{lrrrrrrrr}
    $n_z$ &
    \multicolumn{1}{c}{2} &
    \multicolumn{1}{c}{3} &
    \multicolumn{1}{c}{4} &
    \multicolumn{1}{c}{5} &
    \multicolumn{1}{c}{6} &
    \multicolumn{1}{c}{7} &
    \multicolumn{1}{c}{8} &
    \multicolumn{1}{c}{9} \\

    &
    \multicolumn{1}{c}{\includegraphics[height=0.53\textheight]{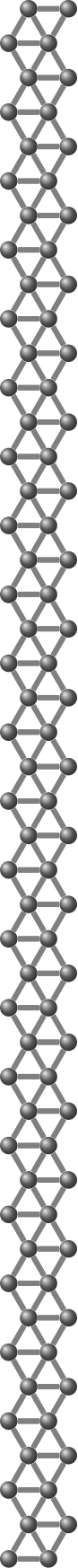}} &
    \multicolumn{1}{c}{\includegraphics[height=0.53\textheight]{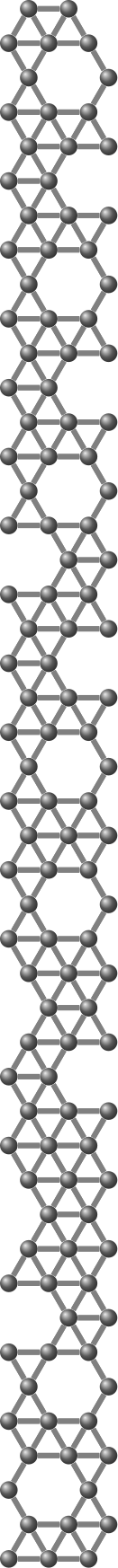}} &
    \multicolumn{1}{c}{\includegraphics[height=0.53\textheight]{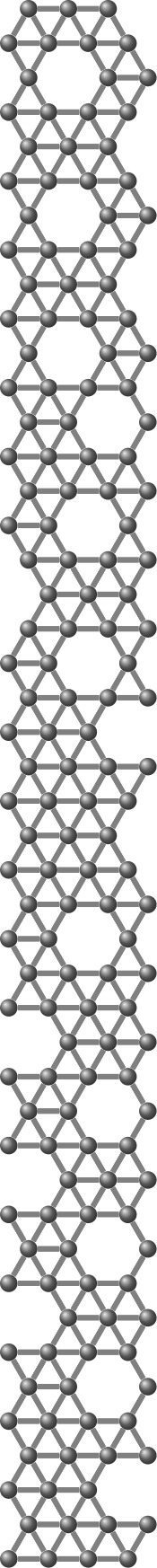}} &
    \multicolumn{1}{c}{\includegraphics[height=0.53\textheight]{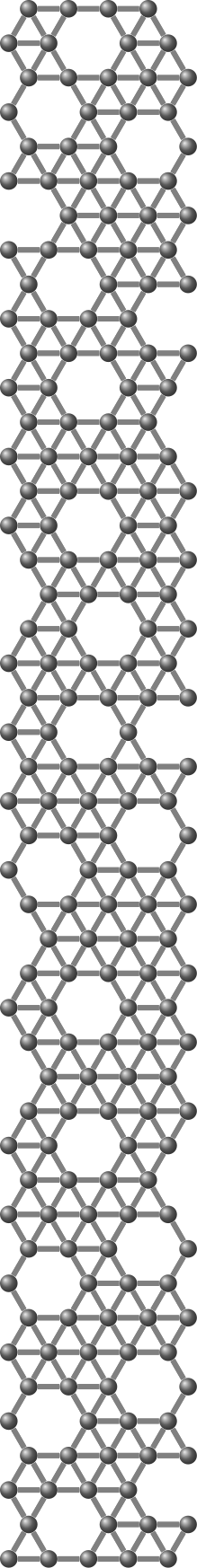}} &
    \multicolumn{1}{c}{\includegraphics[height=0.53\textheight]{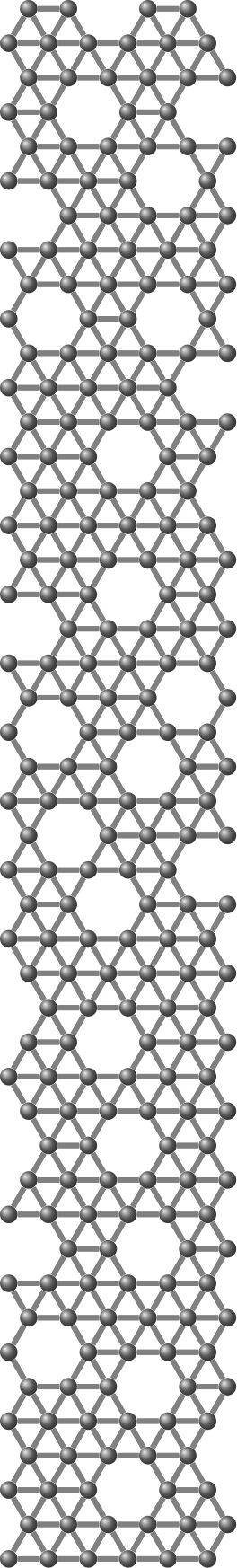}} &
    \multicolumn{1}{c}{\includegraphics[height=0.53\textheight]{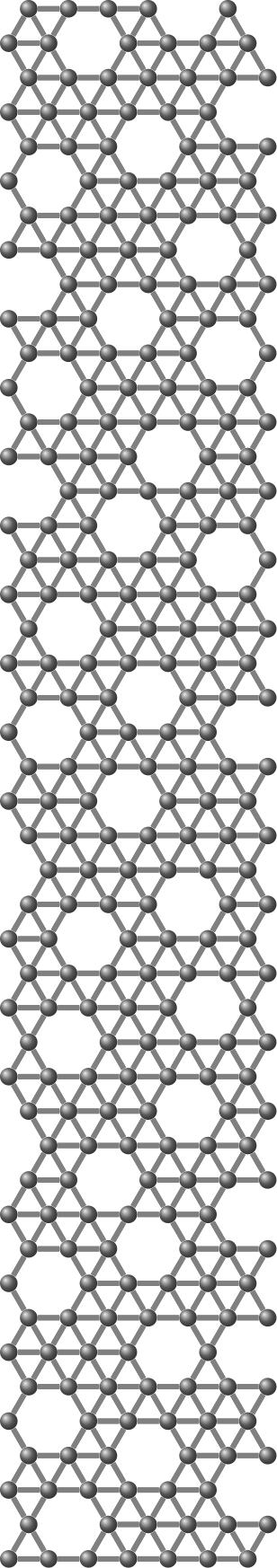}} &
    \multicolumn{1}{c}{\includegraphics[height=0.53\textheight]{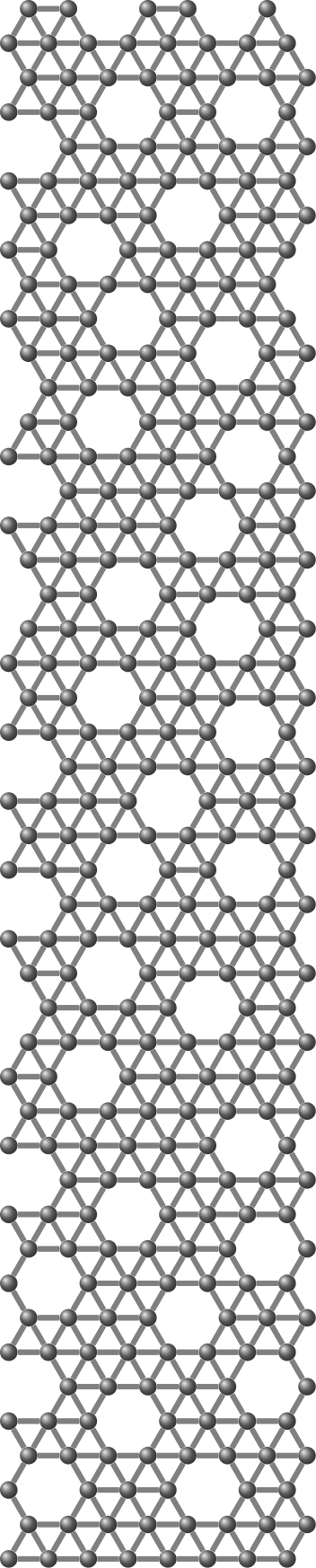}} &
    \multicolumn{1}{c}{\includegraphics[height=0.53\textheight]{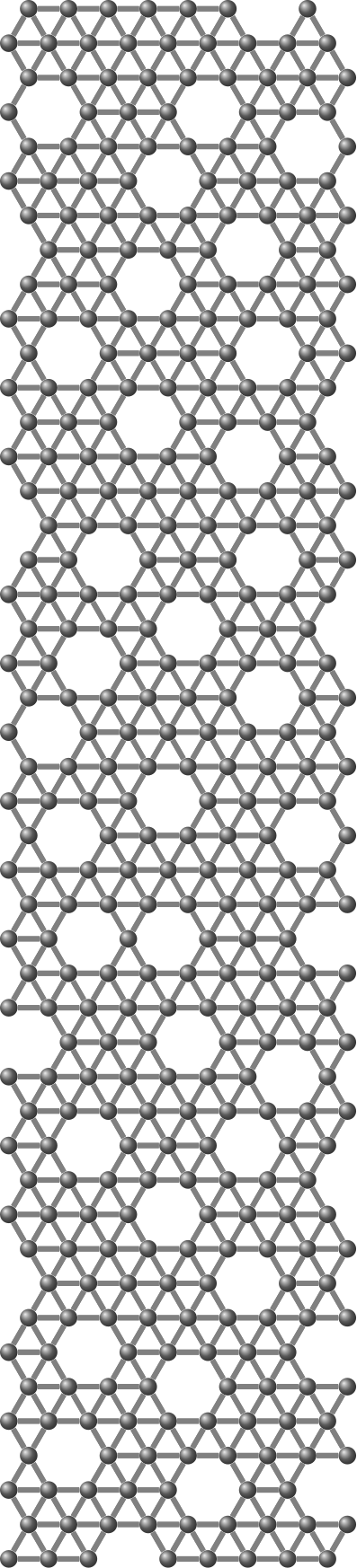}} \\

    &
    $\tau_{23, 2}$ &
    $\tau_{23, 3}$ &
    $\tau_{23, 4}$ &
    $\tau_{23, 5}$ &
    $\tau_{23, 6}$ &
    $\tau_{23, 7}$ &
    $\tau_{23, 8}$ &
    $\tau_{23, 9}$ \\

    $E_{\rm b}$ &
    $6{.}512$ & 
    $6{.}547$ & 
    $6{.}551$ & 
    $6{.}550$ & 
    $6{.}548$ & 
    $6{.}550$ & 
    $6{.}551$ & 
    $6{.}548$ \\

    $c$ &
    92 &
    138 &
    184 &
    230 &
    276 &
    322 &
    368 &
    414 \\

    $N_{\rm a}$ &
    92 &
    122 &
    164 &
    203 &
    247 &
    285 &
    327 &
    370 \\

    $N_{\rm v}$ &
    0 &
    16 &
    20 &
    27 &
    29 &
    37 &
    41 &
    44 \\

    $n_4$ &
    0 &
    4 &
    2 &
    5 &
    4 &
    6 &
    5 &
    6 \\

    $n_5$ &
    0 &
    88 &
    116 &
    152 &
    166 &
    210 &
    236 &
    252 \\

    $n_6$ &
    92 &
    30 &
    46 &
    46 &
    77 &
    69 &
    86 &
    112 \\

    $d_{n_\phi}$ &
    23 &
    1 &
    1 &
    1 &
    1 &
    1 &
    1 &
    1 \\

    $d_{n_z}$ &
    2 &
    1 &
    1 &
    1 &
    1 &
    1 &
    1 &
    1 \\

    $\Delta t$ &    
    2:20:42:33 &
    0:04:43:57 &
    0:03:32:40 &
    0:05:33:33 &
    0:11:15:37 &
    1:00:12:09 &
    2:13:54:34 &
    6:12:22:56 \\
  \end{tabular}
  \endgroup

  \caption{Fragments of boron sheets that if rolled into nanotubes give the most stable 
 BNTs that were obtained in our evolutionary computations. The unit cell of each BNT is defined by two parameters $(n_\phi, n_z)$, where $n_\phi = 23$ and $n_z$ is specified above each stripe. Bellow each stripe are listed several parameters that characterize each nanotube (see text). The binding energy per atom, $E_{\rm b}$, is given in units of ${\rm eV} / {\rm atom}$. The holes in the nanotubes are vacancies in the hexagonal structure wrapped on a cylinder to form a nanotube. The density of holes can be calculated using the formula $\eta = N_{\rm v} / c$, where $c=N_{\rm a}+N_{\rm v}$. The values of the calculation time, $\Delta t$, are presented using the days:hours:minutes:seconds format. One evolutionary process was performed on one logical thread of a multithreaded CPU \cite{CPU}. The unit cell of the $\tau_{23, 8}$ BNT is shown in the inset of Fig.~\ref{fig8}.}
  \label{fig7}
\end{figure}

\begin{figure}
  \centering
  \includegraphics[scale=1.25]{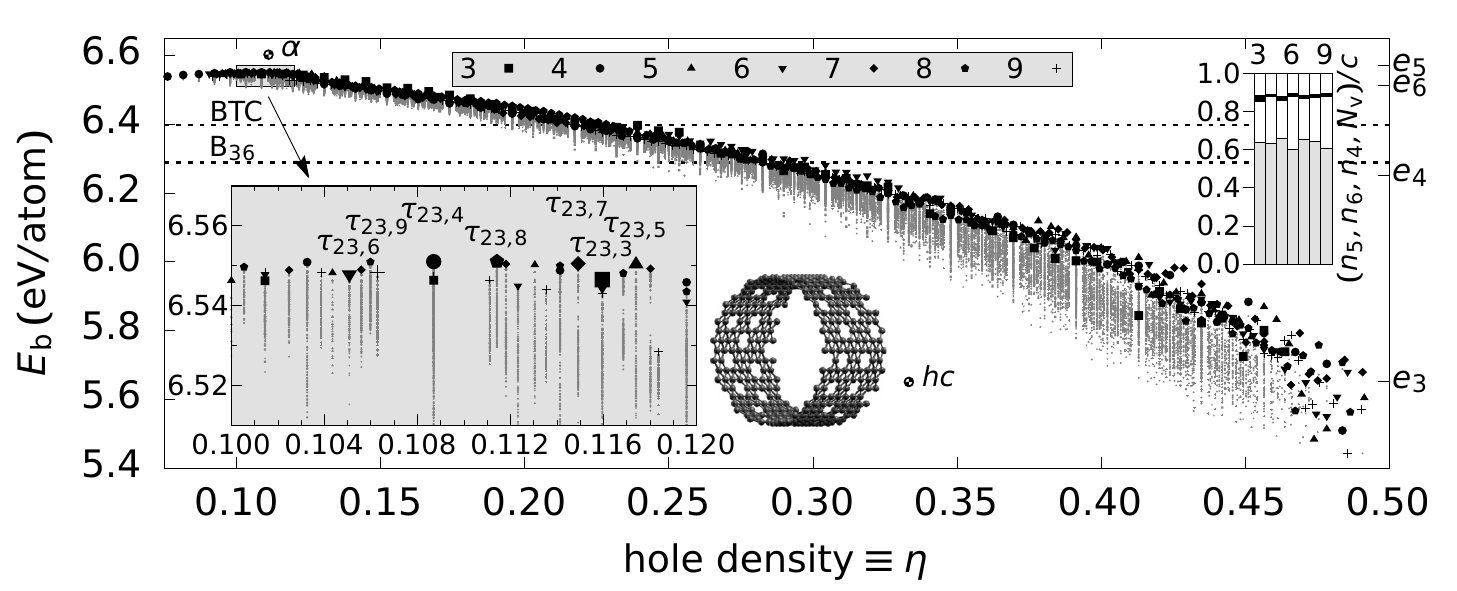}
  \caption{Binding energy per atom, $E_{\rm b}$, as a function of hole density, $\eta$. The data points in black correspond to BNTs with maximum values of $E_{\rm b}$ obtained for a given $n_z$ parameter ($n_\phi = 23$) and shown with a different label. The data points in gray show the full evolution and correspond to nonoptimal BNTs. The $n_z \in \{ 3, 4, \dots , 9\}$ to label correspondence is provided above the data points. For comparison purposes, shown are also the values of $E_{\rm b}$ for the $hc$ and $\alpha$ sheets (with $E_{\rm b}$ equal to $e_3$ and $6{.}603\,{\rm eV/atom}$, respectively \cite{Tarkowski2021}) as well as $E_{\rm b}$ values taken from Ref.~\cite{Tarkowski2021} corresponding to the BTC  and ${\rm B}_{36}$ structures. The inset on the left (with gray background) shows an enlarged view of a fragment of the main picture. The histogram on the right presents (from bottom) the values of $n_5 / c$ (gray), $n_6 / c$ (white), $n_4 / c$ (black), and $N_{\rm v} / c$ (white) for $\tau_{23, n_z}$ nanotubes. In the center, the unit cell of the $\tau_{23, 8}$ nanotube is shown.}
  \label{fig8}
\end{figure}

The most energetically favorable nanotubes are presented in Fig.~\ref{fig7}. The evolutionary obtained structure $\tau_{23, 2}$ is the case of a fully filed with boron atoms lattice. The removal of any atom from a fully filled lattice is energetically favorable since raises the number of 5-coordinated atoms (raise of the beneficial value of $n_5$). The density of vacancies can not be too high since a situation with holes that are too close to each other may be detrimental for $E_{\rm b}$ since may increase the number of 4-coordinated atoms (raise of the detrimental value of $n_4$). Looking at Fig.~\ref{fig7}, we may conclude that, in general, the most energetically favorable are those nanotubes for which the boron holes are separated by fragments of the boron double chain (BDC) or boron triple chain (BTC). Less favorable are those situations in which the holes are next to each other. The highest $E_{\rm b}$ in our simulation belongs to the $\tau_{23, 8}$ nanotube ($6{.}55108\,{\rm eV} / {\rm atom}$) after which the $\tau_{23, 4}$ nanotube ($6{.}55102\, {\rm eV} / {\rm atom}$) is located. Recalling that DFT calculations precision is of the order of $1\,{\rm meV} / {\rm atom}$ \cite{Hoffmann2008}, one can assume that both nanotubes have equal $E_{\rm b}$ (within our model). In Fig.~\ref{fig7}, we also show the time taken to obtain within our approach, the most stable examples of BNTs for a given value of $n_z$. Although our methodology is relatively simple, the time taken to obtain for instance $\tau_{23, 9}$ exceeds 6 days. This tells us that computations that would involve more accurate fitness functions may require prohibitive computational times.

The relation between $E_{\rm b}$ and hole density, $\eta = N_{\rm v} / c$, is presented in Fig.~\ref{fig8}. If one were to draw a curve that is the average of the $E_{\rm b}$ values shown in Fig.~\ref{fig8} as a function of hole density, the maximum of this curve would correspond to the hole density of $1/9 = 0.111$ which is the hole density of the $\alpha$-sheet \cite{Tang2007} (the optimal flat neutral boron 2D structure). The histogram shown in the inset of Fig.~\ref{fig8} presents the number of 4, 5, and 6 coordinated boron atoms as well as boron vacancies relative to the total number of possible boron sites ($c=N_{\rm a}+N_{\rm v}$) for the most stable BNTs. The average over the 7 relevant cases studied ($n_z \in \{ 3, 4, \dots , 9\}$) gives us that the number of 4-, 5-, and 6-coordinated boron atoms is about $2\%$, $63\%$, and $24\%$, respectively, of the total number of atoms forming the nanotubes, whereas about $11\%$ are boron vacancies. This gives us a very interesting result that more than $85\%$ of the boron atoms in BNTs with the largest $E_{\rm b}$ values are highly coordinated atoms.

\begin{figure}
  \centering
  \includegraphics{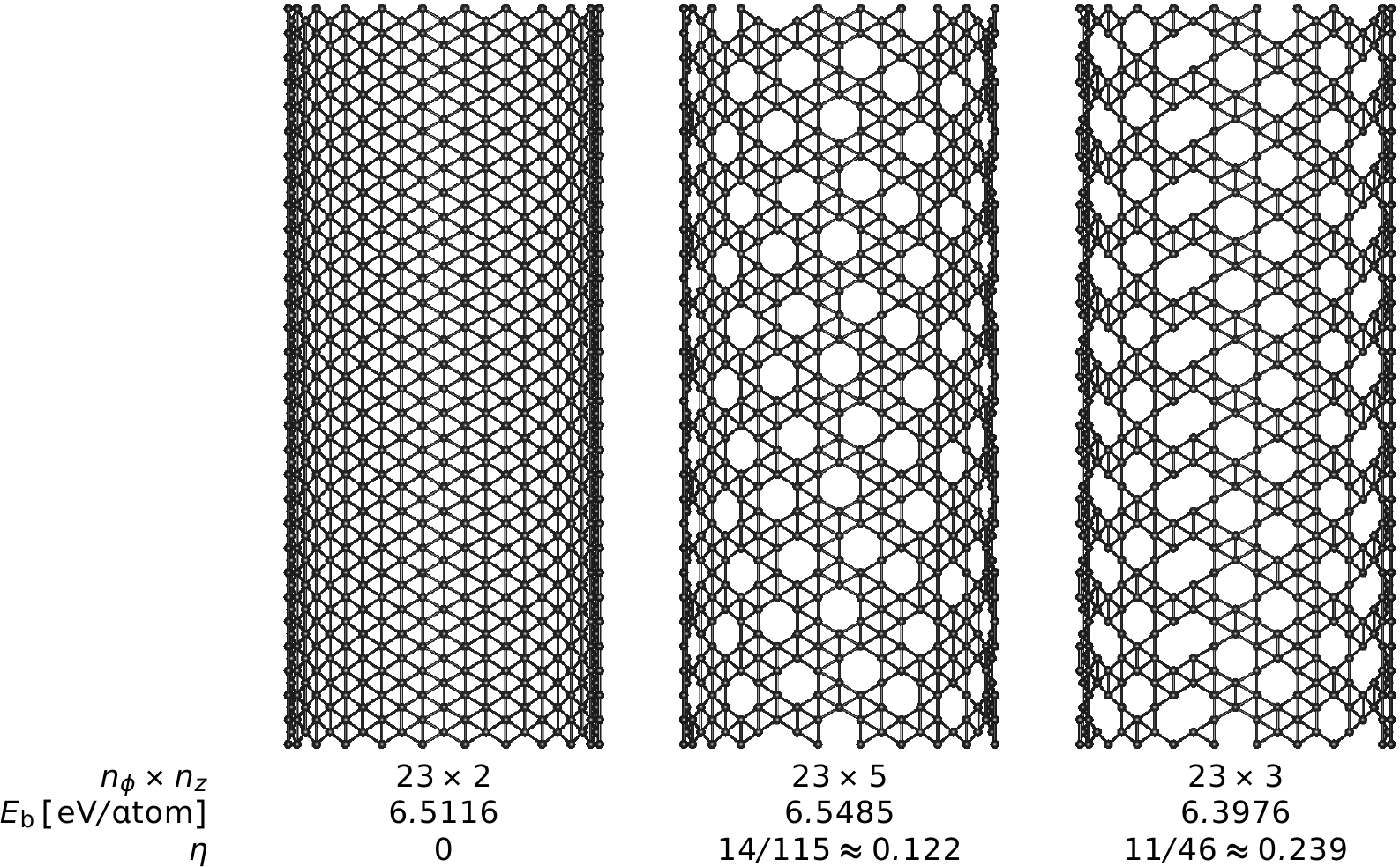}
  \caption{Examples of BNTs with three different hole densities. For each nanotube, the $n_\phi$ and $n_z$ parameters are provided. To obtain the same height of the nanotubes the 2, 5, and 3 values of $n_z$ are multiplied by 15, 6, and 10, respectively (for a total height of 50.25~{\AA}). The binding energy, $E_{\rm b}$, and hole density, $\eta$, is also given.}
  \label{fig9}
\end{figure}

In Fig.~\ref{fig9}, we show three examples of BNTs for three substantially different hole densities $\eta$ equal to 0, 0.122, and 0.239. For hole densities smaller than about 0.22, the boron nanotubes exhibit randomly distributed hexagonal holes and are more stable than a flat stripe structure and a quasi-flat B$_{36}$ cluster (see Fig.~\ref{fig8}). For larger hole densities ($> 0.22$) the boron nanotubes resemble porous tubular structures with hole sizes that depend on the surface densities of boron atoms. Finally should be noted that almost all the studied BNTs in this work have  $E_{\rm b}$ values between $5.65~{\rm eV} / {\rm atom}$ and $6.60~{\rm eV} / {\rm atom}$ which correspond to $E_{\rm b}$ of the $hc$ and $\alpha$ sheets, respectively \cite{Tarkowski2021}.

\section{Summary}
In conclusion, we have presented an adaptation of the genetic algorithm for the structure prediction of tubular BNTs in a broad range of possible hole densities. The evolution of the structures is done using a binary representation and a fitness function calculated using a model Hamiltonian. For the most stable BNTs, the number of 5-coordinated boron atoms is about $63\%$ of the total number of atoms forming the nanotubes, whereas about $11\%$ are boron vacancies. Moreover, more than $85\%$ of the boron atoms in BNTs with the largest $E_{\rm b}$ values are 5- and 6-coordinated atoms. For hole densities smaller than about 0.22, the BNTs exhibit randomly distributed hexagonal holes and are more stable than the BTC structure and a quasi-flat B$_{36}$ cluster. For larger hole densities ($> 0.22$) the BNTs resemble porous tubular structures with hole sizes that depend on the surface densities of boron atoms. These prototype genetic algorithm calculations may serve for the exploration of the structure of nanotubes based on a mixture of boron and other atoms (e.g. carbon and/or nitrogen).

\appendix
\section*{Appendix A}
\addcontentsline{toc}{section}{Appendix A}
Let us define some helper functions describing the indices of six lattice sites (atoms and vacancies) that are neighbors of a given site of the wrapped hexagonal lattice. If a given lattice site has an index $i$, then the sites up left, up right, left, right, down left, and down right have indices equal to $u_l^*(i)$, $u_r^*(i)$, $l^*(i)$, $r^*(i)$, $d_l^*(i)$, and $d_r^*(i)$, respectively. Those functions are formulated in the following way:
\begin{align}
u_l^* & = l^* \circ u_r^*, \\
u_r^*(i) & = (i + n_z + o^*(i)) \bmod 2 n_\phi n_z, \\
l^*(i) & = i - 1 + \left\{ \begin{array}{ll}
  0 & {\rm if}\ i \bmod n_z \neq 0 \\
  n_z & {\rm otherwise}
\end{array} \right., \\
r^*(i) & = i + 1 - \left\{ \begin{array}{ll}
  0 & {\rm if}\ i \bmod n_z \neq n_z - 1 \\
  n_z & {\rm otherwise}
\end{array} \right., \\
d_l^* & = l^* \circ d_r^*, \\
d_r^*(i) & = (i - n_z + o^*(i)) \bmod 2 n_\phi n_z,
\end{align}
where the offset $o^*$ is defined as:
\begin{equation}
o^*(i) = \left\{ \begin{array}{ll}
  0 & {\rm if}\ \lfloor i / n_z \rfloor \ {\rm is\ even} \\
  1 & {\rm if}\ \lfloor i / n_z \rfloor \ {\rm is\ odd} \wedge i \bmod n_z \neq n_z - 1 \\
  1 - n_z & {\rm otherwise}
\end{array} \right.
.\end{equation}
Let ${\rm id}(i) = i$. The following relations hold:
\begin{align}
  u_l^* & = {d_r^*}^{-1}, \\
  l^* & = {r^*}^{-1}, \\
  d_l^* & = {u_r^*}^{-1}, \\
  {u_l^*}^{2n_\phi} & = {d_l^*}^{2n_\phi} = {l^*}^2, \\
  {l^*}^{n_z} & = {r^*}^{n_z} = {\rm id}, \\
  {u_r^*}^{2n_\phi} & = {d_r^*}^{2n_\phi} = {r^*}^2.  
\end{align}
Functions $u_l^*$, $u_r^*$, $r^*$, and $d_r^*$ are used to ensure neighborhood along the nanotube axis direction and at its circumference. To ensure connectivity, an additional set of functions is needed. These functions do not reflect PBCs. If a given atom or vacancy does not possess a suitable neighbor (i.e., it lies at the unit cell boundary), then the appropriate function returns an unchanged index:
\begin{align}
u_l(i) & = i + \left\{ \begin{array}{ll}
  0 & {\rm if}\ \lfloor i / n_z \rfloor = 2n_\phi - 1 \\
  0 & {\rm if}\ \lfloor i / n_z \rfloor \ {\rm is\ even} \wedge i \bmod n_z = 0 \\
  n_z - 1\hphantom{() 0} & {\rm if}\ \lfloor i / n_z \rfloor \ {\rm is\ even} \wedge i \bmod n_z \neq 0 \\
  n_z & {\rm otherwise}
\end{array} \right. ,\\
u_r(i) & = i + \left\{ \begin{array}{ll}
  0 & {\rm if}\ \lfloor i / n_z \rfloor = 2n_\phi - 1 \\
  0 & {\rm if}\ \lfloor i / n_z \rfloor \ {\rm odd} \wedge i \bmod n_z = n_z - 1 \\
  n_z + 1\hphantom{() 0} & {\rm if}\ \lfloor i / n_z \rfloor \ {\rm odd} \wedge i \bmod n_z \neq n_z - 1 \\
  n_z & {\rm otherwise}
\end{array} \right. ,\\
l(i) & = i - \left\{ \begin{array}{ll}
  1 & {\rm if}\ i \bmod n_z \neq 0 \\
  0 \hphantom{(n_z + 1)} & {\rm otherwise}
\end{array} \right. ,\\
r(i) & = i + \left\{ \begin{array}{ll}
  1 & {\rm if}\ i \bmod n_z \neq n_z - 1 \\
  0 \hphantom{(n_z + 1)} & {\rm otherwise}
\end{array} \right. ,\\
d_l(i) & = i - \left\{ \begin{array}{ll}
  0 & {\rm if}\ \lfloor i / n_z \rfloor = 0 \\
  0 & {\rm if}\ \lfloor i / n_z \rfloor \ {\rm is\ even} \wedge i \bmod n_z = 0 \\
  (n_z + 1) \hphantom{0} & {\rm if}\ \lfloor i / n_z \rfloor \ {\rm is\ even} \wedge i \bmod n_z \neq 0 \\
  n_z & {\rm otherwise}
\end{array} \right. ,\\
d_r(i) & = i - \left\{ \begin{array}{ll}
  0 & {\rm if}\ \lfloor i / n_z \rfloor = 0 \\
  0 & {\rm if}\ \lfloor i / n_z \rfloor \ {\rm is\ odd} \wedge i \bmod n_z = n_z - 1 \\
  (n_z - 1) \hphantom{0} & {\rm if}\ \lfloor i / n_z \rfloor \ {\rm is\ odd} \wedge i \bmod n_z \neq n_z - 1 \\
  n_z & {\rm otherwise}
\end{array} \right. .
\end{align}

To formulate the $Q$ predicate, elements of graph theory is used. The set of \textit{vertices}, $V$, consists of indices of all atoms taken from a single unit cell:
\begin{equation}
  V = \left\{ i \in \iota_{2 n_\phi n_z} \mid x_i \right\},
\end{equation}
while the set of (undirected) \textit{edges}, $E$, which vertices fulfill  the neighborhood condition in the nanotube's fragment (unit cell) without PBCs, is now defined as:
\begin{equation}
  E = \left\{ \left\{ i, j \right\} \subseteq V \mid j \in \left\{ u_l(i), u_r(i), l(i), r(i), d_l(i), d_r(i) \right\} \setminus \left\{ i \right\} \right\}.
\end{equation}
By taking a pair of both objects one obtains \textit{graph} $H = \left(V, E\right)$. The $Q$ predicate is defined as $Q \equiv Q_0 \wedge Q_1 \wedge Q_2$, where:
\begin{align}
  Q_0 \equiv & \, \omega (H) = 1, \\
  Q_1 \equiv & \left( \exists i \in \iota_{2 n_\phi n_z} \colon \lfloor i / n_z \rfloor \ {\rm is\ odd} \wedge i \bmod n_z = n_z - 1 \wedge x_i \wedge (x_{u_r^* (i)} \vee x_{d_r^* (i)}) \right) \vee \nonumber \\
  & \left( \exists i \in \iota_{2 n_\phi n_z} \colon \lfloor i / n_z \rfloor \ {\rm is\ even} \wedge i \bmod n_z = n_z - 1 \wedge x_i \wedge x_{r^* (i)} \right), \\
  Q_2 \equiv & \left( \exists i \in \iota_{2 n_\phi n_z} \colon \lfloor i / n_z \rfloor = 2 n_\phi - 1 \wedge x_i \wedge (x_{u_l^* (i)} \vee x_{u_r^* (i)}) \right),
\end{align}
and $\omega (H)$ is the number of \textit{connected components} of graph $H$.

\section*{Acknowledgments}
\addcontentsline{toc}{section}{Acknowledgments}
This work is a result of the projects funded by the National Science Centre of Poland (Twardowskiego 16, PL-30312 Krak\'ow, Poland, \url{http://www.ncn.gov.pl/}) under grants number \mbox{UMO-2013}\allowbreak{}\mbox{/11}\allowbreak{}\mbox{/B}\allowbreak{}\mbox{/ST3}\allowbreak{}\mbox{/04273} and \mbox{UMO-2016}\allowbreak{}\mbox{/23}\allowbreak{}\mbox{/B}\allowbreak{}\mbox{/ST3}\allowbreak{}\mbox{/03575}.

\bibliographystyle{elsarticle-num}
\addcontentsline{toc}{section}{References}
\bibliography{v2}

\end{document}